\documentstyle[prd,epsf,aps,preprint]{revtex}

\newcommand{\lsim}{\lower .5ex\hbox{$\buildrel < \over {\sim}$}}

\def\aa{{\it Astron. Astrophys.}\ }
\def\aap{{\it Astron. Astrophys.}\ }
\def\apj{{\it Astrophys. J.}\ }

\def\apjl{{\it Astrophys.~J.~Letters\ }}

\def\pasp{{\it Publications of the Astronomical Society of the Pacific}\ }
\def\teff{T$_{\rm{eff}}$}

\def\etal{{et~al.\,}}

\def\mj{M$_{\rm J}\,$}
\def\rj{R$_{\rm J}\,$}
\def\mp{M$_{\rm p}$}
\def\rp{R$_{\rm p}$}
\def\rs{R$_{\ast}$}

\def\mstar{M$_{\ast}$}
\def\Dwa{$\,$\uppercase\expandafter{\romannumeral5}$\,$}
\def\mic{$\mu$m$\,$}

\def\sless{\lower2pt\hbox{$\buildrel {\scriptstyle <}
   \over {\scriptstyle\sim}$}}
\def\sgreat{\lower2pt\hbox{$\buildrel {\scriptstyle >}
   \over {\scriptstyle\sim}$}}

\begin{document}


\title{A theoretical look at the direct detection of giant planets outside the Solar
System}
\author{Adam Burrows}
\address{Department of Astronomy, The University of Arizona, Tucson, AZ 85721
\\ e-mail: aburrows@as.arizona.edu}
\maketitle
\setcounter{page}{1}




\section{Heading}
\label{heading}

{\bf

Astronomy is at times a science of unexpected discovery. When
it is, and if we are lucky, new intellectual territories
emerge to challenge our views of the cosmos.  The recent
indirect detections using high-precision Doppler spectroscopy of
now more than one hundred giant planets orbiting more than one
hundred nearby stars is an example of such rare serendipity.
What has been learned has shaken our preconceptions, for none
of the planetary systems discovered to date is like our own.
However, the key to unlocking a planet's chemical, structural, 
and evolutionary secrets is the direct detection of the planet's light.  I review
the embryonic theory of the spectra, atmospheres, and light curves of irradiated  
giant planets and put this theory into the context of
the many proposed astronomical campaigns to image them.  
}

\section{Introduction: The Newly-Discovered Worlds}
\label{sumegp}

Direct detection of an extrasolar planet requires that 
its dim light be separated from 
under the glare of its bright parent star.
However, such high-contrast imaging (e.g., a part in $10^{7-10}$ in the visible) has 
not to date been achieved.   Instead, the vast majority of 
known extrasolar giant planets (EGPs) have been discovered
from the ground using the indirect technique 
of high-precision stellar spectroscopy\cite{MayorQueloz95,mb96,encycl}. 
Due to gravitational attraction, an orbiting planet induces a Doppler wobble in its
parent star.  If the planet is massive and close enough, the periodic
variation in the stellar spectral lines can be measured.  The planet's period ($P$),
eccentricity ($e$), orbital semi-major axis ($a$), and projected mass (M$_{\rm p}\sin(i)$),
where $i$ is the inclination of the orbit, can thereby be determined.  The larger M$_{\rm p}\sin(i)$,
the larger the signal.  This is the reason the first planets
detected were the EGPs.  Terrestrial planets, such as Earth and Venus, are $\sim$300 times
lighter than Jupiter, while ice giants, such as Uranus and Neptune, are $\sim$20 times
lighter.

Before I delve into the physical theory of EGPs 
and their direct detection, I summarize the basic facts
of the known members of the EGP family.   
The first extrasolar giant planet culled was 
51 Peg b \cite{MayorQueloz95} and it is in a tight 4.2-day orbit,
one hundred times closer to its primary than is Jupiter to the Sun.
To date, more than 140 EGPs/planets have been discovered, 
more than 25 of which are in more than 10 multiple systems.  
55 Cancri houses a quadruplet \cite{mcarthur}, one of which 
has a mass near that of Neptune ($\sim$17 Earth masses), 
$\upsilon$ And house a triplet, and GJ 876 houses 
a doublet in a two-to-one orbital resonance. (We follow the convention
by which the planet's name is given by the star's name, with an appended
lower-case letter, either b, c, or d, in discovery order.)

The projected masses of the known Doppler planets vary 
from $\sim$0.06 (!) \mj to above 10 \mj, where \mj
is a Jupiter mass, which is 318 Earth masses
or roughly $10^{-3}$ solar masses.  The more massive objects may be 
brown dwarfs with a different provenance (see Box).  Radial-velocity (Doppler)
techniques can not distinguish EGPs and Neptune-mass planets from brown dwarfs. 
The orbital periods of the known EGPs span a vast range from $\sim$1.2 {\it days} to $\sim$12 {\it years},
their semi-major axes extend from $\sim$0.022 AU to $\sim$6.0 AU, where an AU
is an Astronomical Unit, the distance between the Earth and the Sun,
and their orbital eccentricities vary from 0.0 to above 0.9.
For comparison, Jupiter resides 5.2 AU from our Sun, has an orbital period
of $\sim$12 years, and has an orbital eccentricity of $\sim$0.05.
Table 1 provides these basic data for a representative subset of the current EGP bestiary.
The extremely close-in EGPs, such as 51 Peg b, $\tau$ Boo b, HD209458b,
and OGLE-TR56b\cite{konacki,torres,sasselov03}, were a surprise, but no less so than was
the heterogeneity of the masses and orbital properties of the emerging EGP family.
To be sure, the Doppler technique selects for the closer representatives, but
they must exist to be detected.
As would be expected due to tidal dissipation, the close-in EGPs
with orbital distances smaller than $\sim$0.06 AU all have nearly circular orbits.

There seems to be a correlation between the probability
of finding an EGP and the metallicity of its parent star.  The ``metallicity"
of a star is the mass fraction of elements, such as carbon, oxygen, nitrogen,
neon, magnesium, silicon and iron, that are heavier than helium. 
Hydrogen and helium predominate in stars and giant planets, comprising 
$\sim$98\% by mass of the Sun.  The more super-solar
the heavy-element composition of the potential parent, the more 
likely we are to find an EGP in orbit.  This may be a hint
concerning the processes of giant planet formation, and is
in keeping with the 3-5$\times$solar excesses measured in Jupiter 
and Saturn.  The current census reveals that
there is a $\sim$5\% a priori chance of finding a giant planet by the Doppler technique around 
a nearby (\sless\ 50 parsecs $\equiv$ 160 light-years) star,
but a $\sim$20\% chance of finding one around a star with at least twice the Sun's metallicity
(J. Valenti \& D.A. Fischer, in preparation).

Presumably, the inclinations of EGP orbits are distributed randomly 
on the sky.  Hence, the probability that the orbit is edge-on ($i = 90^{\circ}$) is approximately
\rs/(2$a$), where \rs\ is the stellar radius.   Given this, the
close-in EGPs have the largest chance of transiting the stellar disk,   
during which time the star will dim by a fraction (\rp/\rs)$^2$, where \rp\ is the 
planet's radius.  Since \rj (the radius of Jupiter, $\sim$$7.14\times 10^4$ kilometers) is roughly 10\%
of the radius of the Sun, this ratio is expected to be roughly 1\%.  
A 1\% dimming is easily detectable from the ground.   
At $a$=0.045 AU and a distance ($d$) of 47 parsecs, the planet around the F8V/G0V star HD209458 
was the first of only a handful of EGPs that are now known to
transit their primaries and a periodic dimming at the $\sim$1.6\% level was 
measured \cite{Henry00,Charbonneau00,brown01}.  The transit of HD209458 lasts
$\sim$3 hours (out of a total period of 3.524738 days).  This was followed by the photometrically-selected
transiting EGPs OGLE-TR-56b, OGLE-TR-113b, OGLE-TR-132b, OGLE-TR-111b, 
and TrES-1\cite{konacki,torres,sasselov03,bouchy,alonso,pont}. 
Many more EGP transits are anticipated during the {\it Kepler} 
\cite{Koch98} and {\it Corot} \cite{AntonelloRuiz02} space missions.
These projects are focussed on detecting transits 
around a fraction of the tens of thousands of stars they 
will monitor and will boast photometric accuracy ($\sim$10$^{-5}$) 
sufficient to measure not only transits by EGPs, but by Earth-like planets.
The import of an EGP transit lies in the simultaneous measurement 
of both the orbital inclination (and, hence, with Doppler spectroscopy, the mass) and  
the radius of the planet.  Knowledge of \rp\ and \mp\ (with some knowledge of the star)
can be used to constrain theories of the structure 
and evolution of the close-in EGP \cite{bur.rad,bur.ogle,baraffe}. 
Currently, non-transiting EGPs are mute concerning such physical information.

HD209458 is close and bright enough that the STIS instrument on HST was used 
not only to obtain photometric precision of $\sim$0.01\% \cite{brown01},
but to distinguish a difference at the 4-$\sigma$ 
level in the planetary transit radius in and out of the 
Na-D line at 0.589 \mic.  In this way, neutral sodium
atoms were discovered in HD209458b's atmosphere \cite{Charbonneau02,SeagerSasselov00,Hubbard01}.
Though indirect, this is the first measurement of the composition 
of the atmosphere of an extrasolar planet.  Since then, the Lyman-$\alpha$ 
line of hydrogen has similarly been detected in HD209458b's atmosphere \cite{vidal}, and by
the large magnitude ($\sim$15\%) of the photometric dip 
at this UV wavelength ($\lambda$) a planetary wind\cite{lunine95} comprised
of molecular break-up products has been inferred.   
However interesting, transits are rare and no substitute
for direct imaging and optical and infrared spectra. 
Spectra can provide diagnostics for atmospheric composition, radius, gravity,
and mass.  Images are ground truth for the existence of a planet and 
provide orbital information that complements that gleaned from Doppler
measurements.  Furthermore, direct detection might be able to  
distinguish the different models of giant planet formation, such as nucleation
around an ice/rock core\cite{mizuno} and direct collapse\cite{boss}, and can probe
the outer orbits where the majority of EGPs might reside.   

Since the indirect radial-velocity technique for EGP discovery selects for
the closer variety, it is likely that a large reservoir of giants
exists at distances and orbital periods beyond the reach 
of Doppler spectroscopy.  Furthermore, the best theory
for the orbits of the closest EGPs is that they migrated in from further
out during the early phase of star and planet formation\cite{trilling}.  This too would imply that
a large pool of EGPs resides at larger separations.  Indeed,
it may be that the majority of stars in the solar 
neighborhood harbor planetary systems, that only
new techniques can reveal.  This is where the direct planet detection methods,
most effective at large angular distances from the parent star, 
will come into their own.

\section{Theoretical Atmospheres and Chemistry of EGPs}
\label{chemistry}

After formation, without any significant internal sources of energy, an EGP
gradually cools and shrinks.  Its rate of cooling can be moderated by stellar irradiation,
or by hydrogen/helium phase separation when old and light\cite{fortney04}, but is inexorable.   
Jupiter itself is still cooling and its total infrared
plus optical luminosity is about twice the power intercepted from the Sun.
The rate of cooling is a function of mass and composition, with more massive
EGPs cooling more slowly.  Hence, the instantaneous state of an EGP is a function of mass,
age, composition, orbital distance, and stellar type, not just mass and composition.  

Unlike a star, EGP atmospheric temperatures are sufficiently low that chemistry is destiny.
This is a distinguishing characteristic of substellar-mass objects (SMOs).
The atmosphere of a gaseous giant planet is the thin outer skin of molecules that regulates
its emission spectrum and cooling rate.  Molecular hydrogen (H$_2$) is the overwhelming
constituent, followed by atomic helium.  An EGP's effective 
temperature (\teff, the temperature of its ``photosphere") can vary
from $\sim$1500 K for the more massive EGPs at birth to $\sim$50 K for the least
massive EGPs after a Hubble time.   This wide range translates into a rich variety
of atmospheric constituents that for a given mass and elemental composition evolves
significantly.  At birth, Jupiter had a \teff\ near 600-1000 K and the appearance 
of a T dwarf\cite{burgasser} brown dwarf.  It had no ammonia or water clouds and, due to the presence
of atomic sodium in its hot atmosphere, had a magenta color in the optical\cite{bur.rmp}.  Its atmosphere
was depleted of aluminum, silicon, iron, calcium, and magnesium due to the formation
and settling to depth of the refractory silicates (``dirt") that condense in  
the temperaure range $\sim$1700-2500 K\cite{BurrowsSharp99,Lodders99,Lodders02}.  Water vapor (steam) was the 
major reservoir of oxygen, gaseous methane was the major reservoir of carbon,
gaseous ammonia and molecular nitrogen were the reservoirs
of nitrogen, and H$_2$S was the reservoir of sulfur.   As it cooled, 
the layer of alkali metals was buried below the photosphere to higher pressures, but 
gaseous H$_2$, H$_2$O, NH$_3$, and CH$_4$ persisted to dominate the atmospheric composition.
At a \teff\ of $\sim$400 K, water condensed in the upper atmosphere and water clouds appeared.
This occurred within its first 100 million years.  Within less than a gigayear, when
\teff\ reached $\sim$160 K, ammonia clouds appeared on top of the water clouds,
and this layering persists to this day.  Stellar irradiation retards cloud formation,
as does a large EGP mass, which keeps the EGP hot longer.  
Around a G2V star like the Sun, at 5 Gyr and for an EGP mass of 1.0 \mj,
water clouds form at 1.5 AU, whereas ammonia clouds form beyond 4.5 AU\cite{bur.hub}.  
Jupiter's and Saturn's current effective temperatures are 124.4 K and 95 K, respectively.
Jupiter's orbital distance and age are 5.2 AU and 4.6 Gyr. 
The orbital distance, mass, and radius of a coeval Saturn 
are 9.5 AU, 0.3 \mj, and 0.85 \rj.  However, as an EGP
of whatever mass cools, its atmospheric composition evolves through a similar 
chemical and condensation sequence.  Figure 1 depicts the atmospheric
temperature/pressure (T/P) profile for a sequence of 1-\mj, 5-Gyr models as a function of orbital distance
from a G2V star.  As the planet ``moves" outward, its atmospheric temperature at
a given pressure decreases. Superposed on the plot are the H$_2$O and NH$_3$
condensation lines.  In an approximate sense, a given atmospheric composition 
and temperature can result from many combinations of orbital distance,
planet mass, stellar type, and age.  This lends an added degree of complexity to 
the study of EGPs with which the study of stars does not need to wrestle.

The atmospheres of close-in EGPs (``roasters") at orbital distances of $\sim$0.02-0.07 
AU from a G, F, or K star are heated and maintained at temperatures of 1000-2000 K,
roughly independent of planet mass or composition.  An edge star of the solar-composition, hydrogen-burning
main sequence (\mstar$\sim$75 \mj) has a \teff\ of $\sim$1700 K.
Therefore, an irradiated EGP, with a radius comparable to that of such a star, 
can be as luminous. Its atmospheric composition
is predominantly H$_2$, He, H$_2$O, Na, K, and CO.  At high temperatures, carbon is generally in 
carbon monoxide.  This is the dominant molecule of carbon for M dwarfs with \teff{s} of 2200-3500 K.
At the highest \teff{s}, clouds of iron particulates can form and persist in the upper atmosphere, 
as may be the case in HD209458b.  There are, however, significant day/night
differences and unique reflective properties that distinguish a roaster from 
a lone and isolated edge star.  Exotic general circulation models (GCMs)\cite{menou,Cho03,burk04} may soon be necessary
to understand the equatorial currents, jet streams, day/night differences, terminator
chemistry, and global wind dynamics of severely irradiated roasters, in particular,
and of orbiting, rotating EGPs, in general.  

It is useful to note that a young EGP in a wide orbit with a mass 
of 1.0 to 5.0 \mj has an atmosphere and spectrum that are similar
to those of an old brown dwarf with a mass of 30-60 \mj.  As it evolves, the spectroscopic class of a giant planet can
transition from that of a hot M dwarf, into an L dwarf (where the 
silicate clouds are {\it in} the atmosphere), then into 
a T dwarf, ending up in the territory, as yet unexplored, between the Jovian planets and the ``stars." 
If its mass is low enough, an EGP can cool within gigayears to assume the aspect of our Jovian planets.
Hence, by chemistry, clouds, and \teff, the study of brown dwarfs and EGPs are inextricably linked.

Finally, the best theoretical fits to Saturn's internal structure
suggest that it contains a 5--20 Earth-mass core of heavy elements\cite{saumon}.  This core of (perhaps)
ice and rock may have been the nucleus around which Saturn formed and resembles the ice giants
Neptune and Uranus. The latter may be aborted giant planets that were able to accrete
but little hydrogen from the protosolar/protoplanetary nebula.
The Neptune-mass extrasolar planet, 55 Cancri e, may be
a stripped or aborted EGP.  An alternative mode
of giant planet formation is by direct collapse\cite{boss}.  Under such a scenario, one
would expect a closer correspondence between the heavy-element abundances of
planet and parent star.  Hence, the composition of its atmosphere
and its heavy-element-dependent radius might be keys to an EGP's formation.
These are in principle measureable.

\begin{quote}
\hbox{------------------------------------------------------------------------------------------------------------------}
\centerline{{\bf Box: Brown Dwarfs}}
Brown dwarfs are substellar-mass objects (SMOs)
(\sless\ {0.07} solar masses $\equiv$$\sim$75 \mj) that are
unable to ignite light hydrogen stably to become a
star, but are otherwise formed like stars.  The radiative surface
losses of a star balance the thermonuclear power generated in its core.  This requires
sufficient mass.  The surface losses of a less-massive brown dwarf
are not fully compensated by thermonuclear burning and it cools inexoribly after formation
over a Hubble time.  Nevertheless, brown dwarfs constitute
the low-mass, low-temperature extension of the stellar family
and are an important subject in their own right\cite{bur.rmp,Burrows97}.
Masses in the range of $\sim$10 \mj to $\sim$75 \mj are frequently discussed,
but overlap with the mass distribution of the EGP family is entirely possible.
\end{quote}
\hbox{------------------------------------------------------------------------------------------------------------------}

\section{Spectral Features of EGPs}
\label{spectra}

In principle, as with stellar atmospheres, direct 
detection of the spectrum of an extrasolar giant planet can reveal its
elemental composition, radius, gravity (G\mp/\rp$^2$), and \teff.  Furthermore, when
a cloud dwells in its atmosphere, its associated absorption and scattering properties
might be used to determine the cloud's particle size and makeup. 
Moreover, short-term temporal variations of the planet's flux and spectrum 
might indicate rotation and/or meteorology.  Finally, irradiation introduces
the star-planet-Earth angle as an important parameter, so the orbit's orientation
and instantaneous orbital phase must be factored in (\S\ref{phase}).  Along with 
the dependences on \mp, age, stellar type, and orbital distance, this variety
of influences and parameters makes the study of EGP spectral signatures 
and light curves, and their inversion to obtain planetary properties, rather complicated.

Nevertheless, the molecular mix described in \S\ref{chemistry} determines the emergent
and reflected spectrum.  Though H$_2$ is abundant, it has no permanent electric dipole moment,
and, hence, a very low photon absorption cross section in the optical and infrared.    
Similarly, helium is all but transparent.  The result is that gaseous water vapor,
with its strong absorption features from 0.94 \mic to $\sim$7 \mic, can define much of an 
EGP's spectrum.  Because water resides in both the Earth's and an EGP's atmosphere,
the water bands that bracket and determine the Earth's photometric windows at $\sim$1.0 \mic ($Z$),
$\sim$1.25 \mic ($J$), $\sim$1.65 \mic ($H$), $\sim$2.2 \mic ($K$), $\sim$3.45 \mic ($L^{\prime}$), 
and $\sim$4-5 \mic ($M$), through which ground-based infrared astronomy 
is possible, are exactly the same windows in an EGP or brown dwarf 
atmosphere through which emergent flux can pour.  Thus, and fortuitously
for brown dwarf observations, the emission peaks for SMOs coincide with the
classic Terrestrial atmospheric bandpasses.  

In lieu of measurements, theory fills the vacuum.
Figure 2, taken from 
Burrows, Sudarsky, \& Hubeny\cite{bur.hub}, depicts ``phase-averaged"\cite{Sudarsky00} planet/star flux ratios 
($f$) from 0.5 \mic to 30 \mic for a 1-\mj/5-Gyr EGP in a circular orbit at various distances
from a G2V star like the Sun.  These models are the same as those depicted in Fig. 1.
Similar plots for different assumed parameters
can be generated.  The water absorption troughs are
manifest throughout.  For the closer EGPs at higher atmospheric temperatures,
carbon resides in CO and methane features are weak.  
For these close-in EGPs, the Na-D line at 0.589 \mic and the corresponding resonance
line of K I at 0.77 \mic are important absorbers, suppressing flux in the visible
bands.  Otherwise, the optical flux is buoyed by Rayleigh scattering
of stellar light.   As $a$ increases, methane forms and the methane 
absorption features in the optical (most of the undulations
seen in Fig. 2 for $a$ \sgreat\ 0.5 AU shortward of 1 \mic), at $\sim$3.3 \mic, and at $\sim$7.8 \mic
appear.  Concomitantly, Na and K disappear from the atmosphere
and the fluxes from $\sim$1.5 \mic to $\sim$4 \mic drop.  For all models, the 
mid-infrared fluxes longward of $\sim$4 \mic are due to self-emission, not reflection.
As Fig. 2 makes clear, for larger orbital distances a bifurcation between a 
reflection component in the optical and an emission
component in the mid-infrared appears.  This separation into components
is not so straightforward for the closer, more massive, or younger family members.
For these EGPs, either the large residual heat coming from the core
or the severe insolation prop up the fluxes from 1 to 4 \mic.  The more massive
EGPs, or, for a given mass, the younger EGPs, have larger $J$, $H$, and $K$ band fluxes.
As a result, these bands are diagnostic of mass and age.  For EGPs with large orbital distances,
the wavelength range from 1.5 \mic to 4 \mic between the reflection and emission components
may be the least favorable search space, unless the SMO is massive or young.

When water or ammonia clouds form, scattering off them enhances the optical
fluxes, while absorption by them suppresses fluxes at longer wavelengths in, for example, the 4--5 \mic window.
Because water and ammonia clouds form in the middle of this distance sequence,
the reflection efficiency (or ``albedo"; \S\ref{phase}) is not a monotonic function of $a$.  
These effects are incorporated into Fig. 2, but their precise
magnitude depends upon unknown cloud particle size, composition, and patchiness.
As a consequence, direct spectral measurements might constrain cloud properties.

Importantly, trace non-equilibrium
molecular species, difficult to model, can be present in quantities
sufficient to alter colors.  Such a ``chromophore," 
whose molecular nature is not yet known,
absorbs in the blue and creates the reddish cast of both Jupiter and Saturn,
lowering their albedos shortward of 0.55 \mic by a factor of $\sim$1.5--2.
(Chromophores were not modelled to produce Fig. 2.)

As Fig. 2 suggests, the planet/star contrast ratio is better
in the mid- to far-infrared, particularly at wide separations.  For such separations, the
contrast ratio in the optical can sink to 10$^{-10}$.  For the closest-in EGPs,
such as HD209458b, OGLE-TR56b, 51 Peg b, and $\tau$ Boo b, the contrast
ratio in the optical is between 10$^{-5}$ and 10$^{-6}$ and is more favorable. 
Such EGPs are not shown in Fig. 2; there are in fact about
20 known EGPs with orbital distances less than 0.08 AU.   Due to the
possible formation of iron clouds in their atmospheres, 
HD209458b and OGLE-TR56b may be brighter in the optical than 51 Peg b
and may have higher reflection albedos.  Figure 3 portrays
a generic absolute flux spectrum at 10 parsecs of a close-in EGP (``Class V" in the
nomenclature of Sudarsky et al.\cite{Sudarsky00}),
not unlike HD209458b.  Highlighted are the positions of
some of the important spectral features.   Since modern telescopes can easily
detect fluxes at the milliJansky level, Fig. 3 demonstrates
that the fluxes themselves are not small.  The problem is seeing the planet from under
the glare of the star (\S\ref{telescopes}).  At 10 parsecs and an orbital separation
of 0.05 AU, the maximum angular separation is a challenging 
$\sim$5 milliarcsecs.

\section{Phase Functions for EGPs and Orbital Orientation}
\label{phase}

The planetary albedo and the phases executed 
as planets traverse their orbits are central quantities in the 
theory of EGP light curves.  In addition, as we discuss in this section, 
the wavelength-dependent albedos are strong functions of orbital distance as well. 
Furthermore, the changing orientation  
of the illuminated face of a planet from the Earth's perspective 
translates into a light curve that can show significant flux and color variations.
In Fig. 2, these variations were averaged out over the orbit,
assumed circular. The longitude independence of Jupiter's 
$T/P$ profile results in little day/night variation in the
mid-infrared and, hence, little phase variation, but such a 
planet is too cold to be self-luminous enough in the optical
for its reflected component not to dominate at these shorter wavelengths.  Hence, Jupiter's
optical fluxes can vary from superior conjunction (full face) to 
first quarter (90$^{\circ}$ from superior conjunction, not seen from Earth) or
last quarter (270$^{\circ}$ from superior conjunction, 
also not seen from Earth) by a factor of $\sim$3\cite{dyudina}.
Note that the phase dependence of an EGP's light curve in the mid-infrared
will depend on the degree to which heat can be efficiently redistributed 
over its entire face.  This will depend on 3-dimensional GCM effects that have
not been worked out.  For the close-in EGPs, due to expected day/night temperature
differences\cite{GuillotShowman02,ShowmanGuillot02}, it is likely that there will
be phase variations at all wavelengths.  In particular, phase variations at thermal
wavelengths are likely to shed light on the atmospheric dynamics and longitudinal temperature
distribution of an EGP.

Since its orbit and orientation play such an important role in
an EGP's flux at the Earth and in its interpretation, we summarize
the basic formulae and concepts.  We restrict ourselves to the optical,
for which the concept of an albedo has a clear meaning, but note that
the approach we summarize has general applicability.  

The planet/star flux ratio ($f$) is given by:
\begin{equation}
f = p ({\rm R}_{\rm p}/R)^2 \Phi(\alpha),
\label{falpha}
\end{equation}
where $R$ is the planet/star distance, $p$ is the geometric albedo, $\Phi(\alpha)$ is the phase function,
and $\alpha$ is the star-EGP-Earth angle.  $\Phi(\alpha)$ is normalized to
be 1.0 at full face, thereby defining the geometric albedo, 
and is a decreasing function of $\alpha$.  For so-called ``Lambert"
reflection in which an incident ray on a planetary patch emerges uniformly
over the exit hemisphere, $p$ is 2/3 for purely scattering atmospheres and $\Phi(\alpha)$ is given by the
formula:
\begin{equation}
\Phi(\alpha) = \frac{\sin(\alpha) + (\pi - \alpha)\cos(\alpha)}{\pi}\, .
\label{lambert}
\end{equation}
However, EGP atmospheres are absorbing and the anisotropy of
the single scattering phase function for grains, droplets, or molecules
results in non-Lambertian behavior.  For instance, back-scattering off cloud particles
can introduce an ``opposition" effect for which the planet appears ``anomalously"
bright at small $\alpha$s.  This spike might be a useful signature of cloud particle size.
Moreover, the light scattered from EGPs is likely to be strongly polarized\cite{SWS}.  The degree
of polarization as a function of wavelength and phase angle $\alpha$ can also be used to determine
cloud properties.  However, polarization will be rather more difficult to measure.

Both $p$ and $\Phi(\alpha)$ are functions of wavelength, but the wavelength-dependence
of $p$ is the most severe.  In fact, for cloud-free atmospheres, due
to strong absorption by molecular bands, $p$ can
be as low as 0.03.  Rayleigh scattering serves to support $p$, but mostly in the blue and 
UV, where, however, chromophores can decrease it.  
The presence of clouds increases $p$ significantly.  For instance, at 0.48 \mic, Jupiter's
geometric albedo is $\sim$0.46 and Saturn's is 0.39\cite{kark99}.  Note that for orbital distances
less than 1.5 AU, we expect the atmospheres of most EGPs to be clear.  The albedo
would be correspondingly low.  As a consequence, the 
theoretical albedo is very non-monotonic with distance,
ranging in the visible from perhaps $\sim$0.3 at 0.05 AU, to $\sim$0.05 at 0.2 AU,
to $\sim$0.4 at 4 AU, to $\sim$0.7 at 15 AU \cite{bur.hub,Sudarsky00,Sudarsky03}.  
In the visible ($\sim$0.55 \mic), the geometric 
albedo for a roaster is severely suppressed by Na-D at 0.589 \mic.  Due to a methane
feature, the geometric albedo can vary from 0.05 at $\sim$0.6 \mic to $\sim$0.4
at 0.625 \mic.  Hence, variations with wavelength and with orbital 
distance by factors of 2 to 10 are not unexpected.  Those planning programs of direct 
detection should be aware of such possibilities.

$\Phi(\alpha)$ and $p$ must be calculated or measured, but the sole dependence 
of $\Phi(\alpha)$ on $\alpha$ belies the complications introduced by an orbit's
inclination angle ($i$), eccentricity ($e$), argument of periastron ($\omega$),
and longitude of ascending node ($\Omega$). Along with the period ($P$)
and an arbitrary zero of time, these are the so-called Keplerian elements of an orbit.
Figure 4 diagrams and defines these orientational and orbital parameters.
In the plane of the orbit, the angle between the planet and the periastron/periapse (distance
of closest approach to the star) at the star is $\theta$. In the jargon 
of celestial mechanics, $\theta$ is the so-called ``true anomaly."  For an edge-on orbit ($i = 90^{\circ}$),
and one for which the line of nodes is perpendicular to the line of sight ($\Omega = 90^{\circ}$)
and parallel to the star-periapse line ($\omega = 0^{\circ}$),
$\theta$ is complementary to $\alpha$ ($\alpha = 90^{\circ} - \theta$).  As a result, $\theta = 0^{\circ}$ at 
$\alpha = 90^{\circ}$ (greatest elongation) and increases with time.  Also, for 
such an edge-on orbit, $\alpha = 0^{\circ}$ at superior conjunction.    
In general,
\begin{equation}
\cos(\alpha) = \sin(\theta + \omega)\sin(i)\sin(\Omega) - \cos(\Omega)\cos(\theta + \omega) \, .
\label{cosinef}
\end{equation}
This is merely an application of the law of cosines.

For a circular orbit, $R$ is equal to the semi-major axis ($a$).  However, a planet in an eccentric orbit
can experience significant variation in $R$, and, therefore, 
stellar insolation (by a factor of $(\frac{1+e}{1-e})^2$).  For example, 
if $e = 0.3$, the stellar flux varies by $\sim$3.5 along 
its orbit.  For $e$ = 0.6, this variation
is a factor of 16!  Such eccentricities are by no means rare in the sample of known EGPs (cf. Table 1). 
Therefore, it is possible for the composition of an EGP atmosphere to change significantly
during its orbit, for clouds to appear and disappear, and for there to be delays (``hysteresis")
in the accommodation of a planet's atmosphere to a varying ``insolation" regime.  Ignoring the latter,
eqs. \ref{falpha} and \ref{cosinef} can be combined with $\Phi(\alpha)$ and the standard Keplerian formula
connecting $\theta$ and time for an orbit with a given $P$ and $e$ to derive an EGP's
light curve as a function of wavelength, $i$, $e$, $\Omega$, $\omega$, and time.  
The upshot is that, depending upon orientation and eccentricity, the brightness of an EGP can vary
in its orbit not at all (for a face-on EGP in a circular orbit) or quite dramatically  
(e.g., for highly eccentric orbits at high inclination angles). 
Since astrometric measurements of stellar wobble
induced by EGPs can yield the entire orbit (including inclination), data from
the Space Interferometry Mission (SIM)\cite{UnwinShao00} (expected to achieve
1-microarcsecond narrow-angle accuracy) or Gaia\cite{perryman} could provide important
supplementary data to aid in the interpretation of direct detections of EGPs.


As Saturn itself demonstrates, depending upon orbital orientation, 
planetary rings can greatly augment reflected light
\cite{dyudina,arnold}.  Their possible presence is a 
wild card in the interpretation of direct EGP signatures.
Also, since $\Phi({\alpha})$ is wavelength-dependent,
the potentially large variation in reflected optical flux with epoch 
will be complemented by an interesting variation in color.  The phase functions
$\Phi({\alpha})$ are wavelength-dependent.  For example, planets should
execute trajectories in the color-color space $V-R$ vs. $B-V$, where $B$, $V$, and $R$
are the standard blue, visible, and red bands.  These trajectories will
be functions of cloud particle size, among other 
things, and will be useful atmospheric diagnostics.  Similar behavior 
in the near- and mid-infrared colors, though more modest, may be seen.

\section{Ground-based and Space-based Telescopes for Direct Detection}
\label{telescopes}

As Fig. 2 implies, the wide range of planet/star contrast ratios
and spectral diagnostics suggests different technological solutions to direct
detection.  Furthermore, the relative merits of searching in the optical,
near-infrared, or the mid-infrared have yet to be determined.  Both ground-based
(less expensive) and space-based (more capable) paths are being pursued and while
a discussion that does justice to the many initiatives whose goal
is the remote sensing of EGPs is far beyond the scope of this review, we 
summarize a few representative approaches. 

EGPs, especially if they are young, massive, and close, are bright enough that current
8- to 10-meter class ground-based telescopes or near-term space telescopes (such
as the James Webb Space Telescope (JWST)\cite{mather} with a 6-meter aperture) might be  
sensitive enough to pick up their light.  This is
particularly true in the near- and mid-infrared (see Figs. 2 and 3).   
However, under the extreme glare of its parent star and at small angular separations
of from $\sim$milliarcsecs to around an arcsecond (Table 1), traditional telescope optics
spills far too much light in the vicinity of the planet.  
The major culprits on the ground are the turbulence of the 
atmosphere (``seeing" and scintillation), scattering 
off dust and the spider mount of the secondary, and imperfections in the mirror(s).  Also at issue is
the stability of the optical system.  Even for perfect
optics, the diffraction pattern due to the finite telescope aperture leaves a characteristic
``Airy" pattern that for a ``Jupiter" at 10 parsecs around a solar-type star would
in the optical be hundreds of times brighter.  

Hence, for large (8--10-meter) ground-based telescopes,
(e.g., the two Kecks\cite{Akeson00,AkesonSwain00}, the four 
VLTs\cite{Paresce01,beuzit}, the two Geminis\cite{macintosh}, 
Subaru\cite{tamura}, the binocular LBT\cite{Hinz00,Hinz01})
and mammoth proposed telescopes (e.g., the 100-meter OWL\cite{hawarden}, 
the 20-meter GMT\cite{davison}, the 30-meter
GSMT\cite{strom}), special efforts will be required.
These include adaptive optics (AO) to 
compensate for atmospheric fluctuations (and mirror 
imperfections) with many hundreds or thousands of fast (millisecond) actuators
and very accurate wavefront sensing.  The latter can, in principle, be achieved using artificial laser
guide stars or stars in the field of view.  (With AO, there is usually a bright star
close enough to obviate the need for an artificial beacon.)  Interferometry to null out the stellar light
is also being pursued by the LBT, VLT, and Keck, and the depth of the null is crucial,
as is the angular region over which a sufficient null can be achieved.  Finally, apodizing masks
and/or coronagraphic spots to occult the star and, by diffractive interference, 
redistribute the star's light away from the planet are highly desirable (and may be necessary).
Note that since Dome C in Antartica has some of the best seeing on the planet and the quietest
atmosphere, placing a giant next-generation telescope there may have its advantages\cite{antar}.

An EGP imaging system will be judged by the planet/star contrast ratio, $f$, it 
can achieve at a given wavelength and for a given angular separation from the star.  
Angles of 0.05 to 2.0 arcsecs are contemplated (Table 1), with the requisite contrasts at smaller angles
deemed too difficult for first-generation imaging.  Note 
that the actual requirements are a function of 
distance to the star.  As Fig. 2
indicates for a 5-Gyr/1-\mj EGP at 10 parsecs, $f$s better than $10^{-4}$ at 10 \mic  
and better than $10^{-8}$ in the optical might be necessary.  At the 4-5 \mic bump, $f$s
from $10^{-5}$ to $10^{-8}$ may be called for.  Fortunately, these performance
goals can be relaxed for more massive and younger EGPs at 
angular separations greater than $\sim$0.1$^{\prime\prime}$.  For the 
roasters, almost independent of mass and age, $f$ reaches $10^{-3}$
in the mid-infrared and \sless\ {$10^{-5}$} in the optical, but at 
the corresponding milliarcsecond separations even these contrast ratios
may be too challenging for imaging.  Figure 5 compares the
theoretically required contrast ratios for the fiducial 5-Gyr/1-\mj EGP at various wavelengths
(taken from Fig. 2) as a function of angular separation 
from a solar-type star at 10 parsecs with the putative capabilities
of a sample of proposed imaging systems, both on the ground
and in space.  Contrast ratios for a 0.5-Gyr/7-\mj EGP 
in the $H$ band are also shown.  Orbit and orientation effects  
have been ignored and large error bars should be assigned to both theory and 
projected capability.  In addition, care should be taken to compare theoretical numbers with
experimental hopes for the same wavebands.  In the interests of brevity, we have 
included multiple wavebands on Fig. 5.

Telescopes are stages of components (primary mirror, secondary mirror, lens, 
apertures, apodizing masks, coronagraphs, etc.) that
in series act on the incident source wavefront to focus light of a desired character on instruments.   
Daisy-chained together, each ``optical" component convolves itself with an input wavefront 
to produce an output wavefront.  In spatial frequency space, the operation
of each stage along the optical path is to multiply the Fourier transform
of the incident wavefront by a Fourier transform characteristic 
of that component's optical properties and geometry.   
If you can introduce components in the optical path that filter or alter
the frequency distribution of the wavefront in such a way that its inverse (the
spatial distribution of the light in the last image plane) has little or no light
in a 2-dimensional angular realm around the star where a planet might reside,
then you have a planet-imaging system.  Though a telescope's classical 
angular resolution ($\sim$$\lambda/D$, where $D$ is the diameter of the telescope primary) 
improves with decreasing $\lambda$, the negative effects of the atmosphere actually 
diminish with increasing $\lambda$.  As a result, many ground-based planet-finding 
initiatives (e.g., MMT(AO)\cite{Hinz01}, LBT-I, VLT-I, VLT-PF, Keck-I, 
Gemini-XAOPI/ExAOC, OWL) are planning to optimize 
in the near- or mid-infrared.  Figure 5 summarizes the
performance goals of some of them.    

In coronagraphic mode, the space-bourne JWST may achieve $f$s 
of $10^{-5}$ to $10^{-6}$ for wavelengths from $\sim$1.0 \mic to $\sim$5.0 \mic.
HST/NICMOS has already achieved comparable $f$s in $H$ band 
at angular separations from 0.3$^{\prime\prime}$ to 1.0$^{\prime\prime}$\cite{glenn}.
However, for single space telescopes without the atmosphere with which to contend,
the optical is clearly preferred (small $\lambda/D$).
Curiously, above the atmosphere mirror imperfections and thermal flexure are still problems 
and an AO system is necessary to cancel the wavefront errors introduced
by the corrugations that remain on an otherwise almost perfect mirror surface
after state-of-the-art machining and polishing.  Two major space-based projects
to image EGPs are being proposed.  The first, $EPIC$\cite{shaoepic}, is
a nulling coronagraph which converts a single telescope pupil into a         
multi-beam nulling interferometer, producing a null which 
is then filtered by an array of single-mode fibers to suppress the residual scattered light.
The design goal of $EPIC$ is for $f$s of $10^{-9}$ to 10$^{-10}$.
The second, {\it ECLIPSE}\cite{Trauger00,Trauger01}, is an 
off-axis coronagraph with an exquisitely-figured 1.8-meter 
primary that is designed to achieve $f$s in the $V$ band better than $10^{-9}$
for angular separations from $\sim$0.1$^{\prime\prime}$ to $\sim$2.0$^{\prime\prime}$.  
Both {\it ECLIPSE} and $EPIC$ will be challenging, 
but if successful will directly detect within $\sim$7 years many 
EGPs in the solar neighborhood out to 10--15 
parsecs.     

However, the flagship of the NASA {\it Origins} program, the Terrestrial 
Planet Finder (TPF\cite{beichman}), whose goal is to image planetary systems and extrasolar
Earths and to obtain low-resolution ($\lambda/\Delta\lambda$ = 10-20) spectra that may reveal 
in rudimentary fashion the O$_2$, H$_2$O, CH$_4$, O$_3$, or CO$_2$ signatures
of life, will also be a formidable instrument for directly
detecting and characterizing EGPs.  As Fig. 5 indicates, 
for angular separations between $\sim$0.05$^{\prime\prime}$
and $\sim$2.0$^{\prime\prime}$, either the more-straightforward optical coronagraphic design (TPF-C)
or the multi-telescope infrared (5--20 \mic) interferometer (TPF-I/Darwin)
would detect EGPs much more readily than the extrasolar Earths that are its 
primary targets.  

For the close-in EGPs, direct imaging seems out of the question for the forseeable future.  
However, this does not mean that the planetary flux can not be measured.
From the ground, there are a variety of techniques to use the planet-plus-star
light to distinguish the planetary component (particularly for known roasters).  These include 1) 
using precision photometry to a part in $10^{5}$ (!) to measure the 
phase variations of the summed optical or near-infrared light, 2) measuring the motion
of the light centroid, perhaps best done in the mid-infrared with an Antarctic
30-meter telescope, 3) spectral deconvolution of a known EGP/star system using its RV-measured 
velocities and ephemeris, and 4) multi-frequency differential interferometric imaging (pioneered for Keck,
among others).  Further transit studies of HD209458b (such as led to the discovery of
the Na-D and Lyman-$\alpha$ features) are certainly warranted.  The 
ground-based methods will be challenging, but less expensive
than space-based efforts.  However, there is currently in space a micro-satellite, MOST\cite{Matthews01},  
with a 15-centimeter aperture, that is designed to achieve photometric accuracy
in the optical of a few$\times 10^{-6}$.  It has on its current observing manifest 
programs to stare at 51 Peg b, $\tau$ Boo b,
and HD209458b.  There have as yet been no announcements.

\section{The Future}
\label{conclusions}

With many programs of direct planet detection planned on a large subset of the 
LBT, VLT, Keck, Gemini, GMT, GSMT/CELT, and OWL on the ground and
HST, {\it Corot}, {\it Kepler}, MOST, SIM, Gaia, TPF-C, TPF-I/Darwin, {\it ECLIPSE}, EPIC, and JWST
in space, during the next twenty years there will be an increasing crescendo
of new results on extrasolar planets that will completely transform our view of the
nature of planetary systems.

EGPs, being brighter, are the natural technological and 
scientific stepping stones on the path to imaging 
extrasolar Earths. We will encounter them first.
Both the NASA and ESA roadmaps\cite{origins} have
given planet detection pride of place.  Strategies are now being formulated
to establish a logical sequence of missions and telescope construction that
will optimize the pace of discovery.  Moreover, theoretical work in support
of mission planning is maturing to the point that it may be ready to interpret
what we observe.  However, a theorist's prejudices aside, one can't help but wonder: What
is it we will actually find?  What discoveries will be made?  As the hunt
for worlds beyond our solar system quickens, an ancient curiosity stirs to ask:
What will {\it our} generation see from that fabled peak in Darien\cite{quote}?


{}

Correspondence and requests for materials should be addressed to Adam Burrows (burrows@as.arizona.edu).


The author acknowledges David Sudarsky, Ivan Hubeny, Bill Hubbard, Jonathan Lunine,
Jim Liebert, Jason Young, John Trauger, Jonathan Fortney, Aigen Li, Christopher Sharp, and Drew Milsom
for fruitful conversations or technical aid and help during the
course of this work, as well as
NASA and the NASA Astrobiology 
Institute for their financial support.  



\begin{table}
 \caption{Interesting EGPs Listed by Angular Separation\label{data.angular}}
\begin{center}
\begin{tabular}{cccccccc} 
{EGP} & $a(1+e)/d$ ($^{\prime\prime})$$^a$ & {star}
& {a (AU)} & {d (pc)}
& {P} & {M$_p$sin($i$) (\mj)} & {$e$}\\
\hline
$\epsilon$ Eri b & 1.61 & K2V & 3.3 & 3.2  & 6.85 yrs. & 0.86 & 0.61 \\
55 Cnc d          & 0.51 & G8V & 5.9 & 13.4 & 14.7 & 4.05 & 0.16  \\
47 UMa c          & 0.31 & G0V& 3.73& 13.3 & 7.10 & 0.76 & 0.1 \\
HD 160691c        & 0.27 & G3IV-V&2.3&15.3& 3.56&$\sim$1&$\sim$0.8\\
$\upsilon$ And d  & 0.27& F8V& 2.50& 13.5 & 3.47 & 4.61 & 0.41 \\
HD 39091b         & 0.26& G1IV& 3.34& 20.6 & 5.70 & 10.3 & 0.62 \\
Gl 777A b         & 0.23& G6V& 3.65& 15.9 & 7.15 & 1.15 & $\sim$0 \\
14 Her b          & 0.20 & K0V& 2.5 & 17   & 4.51 & 3.3  & 0.33 \\
47 UMa b          & 0.17 & G0V& 2.09& 13.3 & 2.98 & 2.54 & 0.06 \\
HD 33636b         & 0.17 & G0V& 3.56 & 28.7 & 4.43 & 7.71 & 0.41 \\
HD 10647b         & 0.16 & F9V& 2.10 & 17.3 & 2.89 & 1.17 & 0.32 \\
$\gamma$ Cephei b & 0.15 & K2V& 1.8& 11.8 & 2.5 & 1.25 & $\sim$0 \\
HD 147513b        & 0.15 & G3V& 1.26& 12.9 & 1.48 & 1.0  & 0.52 \\
HD 216437b        & 0.134 & G4V & 2.7 & 26.5 & 3.54 & 2.1  & 0.34 \\
HD 160691b        & 0.127& G3IV-V&1.48&15.3 & 1.74 & 1.7  & 0.31 \\
HD 70642b         & 0.121& G5IV-V&3.3& 29   & 4.79 & 2.0  & 0.10 \\
HD 50554b         & 0.109 & F8V& 2.38& 31.03& 3.50 & 4.9  & 0.42 \\
HD 106252b        & 0.108 & G0V& 2.61& 37.44& 4.11 & 6.81 & 0.54 \\
HD 168443c        & 0.107 & G5V& 2.87& 33   & 4.76 & 17.1 & 0.23 \\
HD 10697b         & 0.075& G5IV& 2.0 & 30   & 2.99 & 6.59 & 0.12 \\
\\
$\upsilon$ And c  & 0.072 & F8V & 0.83 & 13.5 & 241 days & 2.11 & 0.18 \\
GJ 876b           &0.049 & M4V & 0.21 & 4.72 & 61.0 & 1.89 & 0.1 \\
GJ 876c           &0.036 & M4V & 0.13 & 4.72 & 30.1 & 0.56 & 0.27 \\
HD 114762b        &0.017& F9V&0.35 & 28 & 84.0 & 11.0 & 0.34 \\
55 Cnc b          &$8.4\times 10^{-3}$&G8V& 0.12 & 13.4 & 14.7 & 0.84 & 0.02 \\
$\upsilon$ And b  &$4.5\times 10^{-3}$&F8V& 0.059 & 13.5 & 4.62 & 0.71 & 0.034 \\
51 Peg b          &$3.4\times 10^{-3}$&G2V& 0.05 & 14.7 & 4.23 & 0.44 & 0.01 \\
$\tau$ Boo b      &$3.3\times 10^{-3}$&F7V& 0.05 & 15 & 3.31 & 4.09 & $\sim$0 \\
HD 49674b         &$1.6\times 10^{-3}$& G5V&0.057& 40.7 & 4.95 & 0.12 & 0.17 \\
HD 209458b        &$9.6\times 10^{-4}$&G0V& 0.045& 47 & 3.52 & 0.69 & $\sim$0 \\
HD 83443b         &$9.4\times 10^{-4}$& K0V&0.038& 43.5 & 2.99 & 0.35 & 0.08 \\
OGLE-TR56b        &$1.5\times 10^{-5}$& G0V&0.023& $\sim$1500 & 1.21 & 1.45 & $\sim$0\\
\end{tabular}
\end{center}
$^a$ Maximum possible angular separation (at apoapse/apastron).
\end{table}


{Figure 1.  Profiles of atmospheric temperature (in Kelvin) 
versus the logarithm base ten of the pressure (in bars)
for a family of irradiated 1-\mj EGPs around a G2V star 
as a function of orbital distance.
Note that the pressure is decreasing along the ordinate, which thereby resembles altitude.
The orbits are assumed to be circular, the planets are assumed to have a radius of 1 \rj,
and the orbital separations vary from 0.2 AU to 15 AU.
The intercepts with the dashed lines identified with either \{NH$_3$\} or \{H$_2$O\} denote the
positions where the corresponding clouds form. Taken 
from Burrows, Sudarsky, and Hubeny\cite{bur.hub}.  
See text for a discussion.}

{Figure 2. Planet to star flux ratios versus wavelength (in microns) from 0.5 \mic to 30 \mic
for a 1-\mj EGP with an age of 5 Gyr orbiting a G2V main sequence star similar to the Sun.
This figure portrays ratio spectra as a function of orbital distance from 0.2 AU
to 15 AU.  Zero eccentricity has been assumed and the planet spectra have been phase-averaged
as described in Sudarsky, Burrows, and Hubeny\cite{Sudarsky03}.
The associated $T/P$ profiles are given in Fig. 1. 
Note that the planet/star flux ratio is most favorable in the
mid-infrared.  Water features at 0.94 \mic, 1.2 \mic, 1.4 \mic, 1.9 \mic,
2.6 \mic, and 6.5--8 \mic, methane features in the optical and at 0.89 \mic, 2.2 \mic, 3.3 \mic,
and 7.8 \mic, carbon monoxide features at 2.3 \mic and 4.67 \mic, the Na-D doublet 
at 0.589 \mic, and the K I doublet at 0.77 \mic help shape these spectra.
Taken from Burrows, Sudarsky, and 
Hubeny\cite{bur.hub}.  See text for discussion.}

{Figure 3.  The logarithm of the absolute flux in milliJanskays 
($\equiv 10^{-26}$ ergs cm$^{-2}$ s$^{-1}$ Hz$^{-1}$) at 10 parsecs
for a ``Class V" roaster versus wavelength (in microns) from 0.4 \mic to 5 \mic.
This could be a 1-\mj EGP in a 0.05 AU orbit around a solar-type star.  
The planet spectrum has been phase-averaged
as described in Sudarsky et al.\cite{Sudarsky00,Sudarsky03}.
Shown are the positions of various relevant molecular bands and atomic lines.
Figure taken from Sudarsky, Burrows, \& Hubeny\cite{Sudarsky03}.  
See text for discussion.}

{Figure 4.  Keplerian orbital elements.  The intersection of the orbit plane with the 
observational plane defines the angles $i$, $\omega$, $\Omega$, and $\theta$.
The angle between the observer (Earth) and the line of nodes 
(intersection of the orbit plane with the horizontal plane)
is the longitude of the ascending node ($\Omega$), the angle between the line
of nodes and the focus(star, in yellow)--periapse(black dot) is the argument of periastron
($\omega$), the angle between the orbit plane and the Z-axis 
(perpendicular to the horizontal plane) is the inclination ($i$), and the angle
between the focus--periapse line and the position of the planet (red dot) is
the true anomaly ($\theta$).
See text for details.}

{Figure 5.  A comparison of the planet/star contrast ratios 
(and contrast magnitudes $= -2.5\log(f)$) versus angular separation
(in arcseconds) achievable for some proposed planet imaging systems.  
A distance of 10 parsecs is assumed.
Integration times and signals-to-noise assumed vary and are taken from
preliminary studies by the associated instrument teams.
At $H$ band (red), the imaging telescopes represented include 
the Canada/France/Hawaii Telescope (CFHT)\cite{racine},
HST/NICMOS, and the Gemini/XAOPI.  Not shown on this plot is the 
MMT (AO) system, which should achieve at a wavelength of 5 \mic $f$s from 10$^{-4}$ 
to a few $\times 10^{-6}$ for angular separations of 0.3$^{\prime\prime}$
to 1.0$^{\prime\prime}$, respectively.  The LBT, also not shown,
should achieve at a wavelength of 10 \mic approximately 10$\times$ 
better performance than this. At 5 \mic, a notional curve for a 
20- or 30-meter telescope in Antarctica and the JWST (in fact at 4.6 \mic) are provided. 
At 10 \mic, a notional curve for a 100-meter in Antarctica is given.  
Also included on this plot is the interferometric version 
of TPF (TPF-I/Darwin), which might have a sensitivity  
of one part in $10^7$ from 5 \mic to 20 \mic.  All the mid-infrared
curves are in blue.  In the optical (green, $V$),
putative sensitivities for $EPIC$, {\it ECLIPSE}, and TPF-C are plotted.
Superposed are corresponding ``phase-averaged" 
theoretical curves (dashed) for a 5-Gyr/1-\mj EGP around a G2V star in the $H$ band ($\sim$1.65 \mic),
in the 4--5 \mic band, and at 10 \mic (see Fig. 2).  Also included are a theoretical curve
in the $H$ band (dashed red) for a 0.5-Gyr/7-\mj EGP around a G2V star and a green swathe
where the known EGPs may reside in the optical ($V$ band).
Note that the theoretical curves for more massive 
and younger EGPs than represented on this plot can be {\it considerably} higher.
Orientation effects have been ignored.
Each curve is for a given wavelength or 
bandpass and the imager and theory must be compared at the same wavelength.   
Very generous error bars should be assumed. 
The photometric sensitivity curves for MOST and {\it Kepler} are also superposed.
See text for details.}


\begin{thebibliography}{99}

\bibitem{MayorQueloz95} Mayor, M. \& Queloz, D. 
A Jupiter-Mass Companion to a Solar-Type Star.
{\it Nature} {\bf 378}, 355 (1995).

\bibitem{mb96} Marcy, G.W. and R.P. Butler, R.P.  
A Planetary Companion to 70 Virginis.
\apj {\bf 464}, L147-151 (1996).

\bibitem{encycl} see J. Schneider's
Extrasolar Planet Encyclopaedia at \verb"http://www.obspm.fr/encycl/encycl.html"
for an up-to-date listing.

\bibitem{mcarthur} MacArthur, B. et al. Detection of a Neptune-mass planet
in the $\rho^1$ Cancri system using the Hobby-Eberly Telescope. (55 Cancri e)
submitted to \apj (2004).

\bibitem{konacki} Konacki, M., Torres, G., Jha, S., and Sasselov, D.
An extrasolar planet that transits the disk of its parent star.
{\it Nature}  {\bf 421}, 507-509 (2003).

\bibitem{torres} Torres, G, Konacki, M., Sasselov, D., \& Jha, S. 
New Data and Improved Parameters for the Extrasolar Transiting Planet OGLE-TR-56b.
\apj {\bf 609}, 1071-1075 (2003).

\bibitem{sasselov03} Sasselov, D. 
The New Transiting Planet OGLE-TR-56b: Orbit and Atmosphere.
\apj {\bf 596}, 1327-1331 (2003).


\bibitem{Henry00} Henry, G., Marcy, G.W., Butler, R.P.,
\& Vogt, S.S. A Transiting ``51 Peg-like'' Planet. \apj {\bf 529}, L41-44 (2000).

\bibitem{Charbonneau00} Charbonneau, D.,
Brown, T.M., Latham, D.W., \& Mayor, M. 
Detection of Planetary Transits Across a Sun-like Star.
\apjl {\bf 529}, L45-48 (2000).

\bibitem{brown01} Brown, T. M., Charbonneau, D.,
Gilliland, R.L., Noyes, R.W., and Burrows, A. 
Hubble Space Telescope Time-Series Photometry of the Transiting Planet of HD 209458.
\apj {\bf 552}, 699-709 (2001).

\bibitem{bouchy} Bouchy, F., Pont, F., Santos, F.C., Melo, C.,
Mayor, M., Queloz, D., \& Udry, S. 
Two new ``very hot Jupiters'' among the OGLE transiting candidates.
\aap {\bf 421}, L13-16 (2004).  

\bibitem{alonso} Alonso, R. et al. TrES-1: The transiting planet of a bright K0V star.
in press (2004).

\bibitem{pont} Pont, F., Bouchy, F., Queloz, D., Santos, N.C., Melo, C., Mayor, M., \& Udry, S.
The ``missing link": A 4-day period transiting exoplanet around OGLE-TR-111.
\aap, in press (astro-ph/0408499) (2004).

\bibitem{Koch98} Koch, D., Borucki, W., Webster, L.,
Dunham, E., Jenkins, J., Marrion, J., \& Reitsema, H. 
Kepler: a space mission to detect earth-class exoplanets. {\it SPIE
Conference 3356: Space Telescopes and Instruments V}, 599-607 (1998).

\bibitem{AntonelloRuiz02} Antonello, E. \&
Ruiz, S.M. The Corot Mission.
{\it Memoria della Societa Astronomica Italiana} {\bf 73}, 1241 (2002).

\bibitem{bur.rad} Burrows, A., Sudarsky, D., \& Hubbard, W.B. 
A Theory for the Radius of the Transiting Giant Planet HD 209458b.
\apj {\bf 594}, 545-551 (2003).

\bibitem{bur.ogle} Burrows, A., Hubeny, I., Hubbard, W.B., \& Sudarsky, D. 
Theoretical Radii of Transiting Giant Planets: The Case of OGLE-TR-56b.
\apj {\bf 610}, L53-56 (2004).

\bibitem{baraffe} Baraffe, I., Chabrier, G., Allard, F., and Hauschildt, P.H. 
Evolutionary models for cool brown dwarfs and extrasolar giant planets. The case of HD 209458.
\aa {\bf 402}, 701-712 (2003).

\bibitem{Charbonneau02} Charbonneau, D.,
Brown, T.M., Noyes, R.W., \& Gilliland, R.L. 
Detection of an Extrasolar Planet Atmosphere.
\apj {\bf 568}, 377-384 (2002).

\bibitem{SeagerSasselov00} Seager, S.
\& Sasselov, D.D. Theoretical Transmission Spectra during Extrasolar Giant Planet Transits.
\apj {\bf 537}, 916-921 (2000).

\bibitem{Hubbard01} Hubbard, W.B., Fortney, J.F.,
Lunine, J.I., Burrows, A., Sudarsky, D., \& Pinto, P.A.
Theory of Extrasolar Giant Planet Transits.
\apj {\bf 560}, 413-419 (2001).

\bibitem{vidal} Vidal-Madjar, A., des Etangs, A., Desert, J.-M., Ballester, G.E.,
Ferlet, R., Hebrard, G., \& Mayor, M. 
An extended upper atmosphere around the extrasolar planet HD209458b.
{\it Nature} {\bf 422}, 143-146 (2003).

\bibitem{lunine95} Burrows, A. \& Lunine, J.I. 
Extrasolar Planets - Astronomical Questions of Origin and Survival.
{\it Nature} {\bf 378}, p. 333 (1995).

\bibitem{mizuno} Mizuno, H. Formation of the Giant Planets.
{\it Progress of Theoretical Physics} {\bf 64}, 544-557 (1980).

\bibitem{boss} Boss, A.P. Gas Giant Protoplanet Formation: Disk 
Instability Models with Thermodynamics and Radiative Transfer.
\apj {\bf 563}, 367-373 (2001).

\bibitem{trilling} Trilling, D., Benz, W.,
Guillot, T., W.B. Hubbard, W.B., J.I. Lunine, \& Burrows, A.
Orbital Migration of Giant Planets: Modeling Extrasolar Planets.
\apj {\bf 500}, 428-439 (1998).

\bibitem{fortney04} Fortney, J.J. \& Hubbard, W.B.
Effects of Helium Phase Separation on the Evolution of Extrasolar Giant Planets.
\apj {\bf 608}, 1039-1049 (2004).


\bibitem{burgasser} Burgasser, A. et al. The Spectra of
T Dwarfs. I. Near-Infrared Data and Spectral Classification.
\apj {\bf 564}, 421-451 (2002).

\bibitem{bur.rmp} Burrows, A., Hubbard, W.B., Lunine, J.I.,
\& Liebert, J. The theory of brown dwarfs and extrasolar giant planets.
{\it Rev. Mod. Phys.} {\bf 73}, 719-765 (2001).

\bibitem{BurrowsSharp99} Burrows, A. \&
Sharp, C.M. Chemical Equilibrium Abundances in Brown Dwarf and Extrasolar Giant Planet Atmospheres.
\apj {\bf 512}, 843-863 (1999).

\bibitem{Lodders99} Lodders, K. 
Alkali Element Chemistry in Cool Dwarf Atmospheres.
\apj {\bf 519}, 793-801 (1999).

\bibitem{Lodders02} Lodders, K. \& Fegley, B. 
Atmospheric Chemistry in Giant Planets, Brown Dwarfs, and 
Low-Mass Dwarf Stars. I. Carbon, Nitrogen, and Oxygen.
{\it Icarus} {\bf 155}, 393-424 (2002).

\bibitem{bur.hub} Burrows, A., Sudarsky, D., \& Hubeny, I. 
Spectra and Diagnostics for the Direct Detection of Wide-Separation Extrasolar Giant Planets.
\apj {\bf 609}, 407-416 (2004).

\bibitem{menou} Menou, K., Cho, J. Y-K., Seager, S. \& Hansen, B.M.S.
``Weather'' Variability of Close-in Extrasolar Giant Planets.
\apj {\bf 587}, L113-116 (2003).

\bibitem{Cho03} Cho, J. Y-K., Menou, K., Hansen, B.M.S.,
\& Seager, S. The Changing Face of the Extrasolar Giant Planet HD 209458b.
\apj {\bf 587}, L117-120 (2003).

\bibitem{burk04} Burkert, A., Lin, D.N.C., Bodenheimer, P., Jones, C., \& Yorke, H. 
On the Surface Heating of Synchronously-Spinning Short-Period Jovian Planets.
astro-ph/0312476 (2004).

\bibitem{saumon} Saumon, D. \& Guillot, T. Shock Compression
of Deuterium and the Interiors of Jupiter and Saturn.
\apj {\bf 609}, 1170-1180 (2004).

\bibitem{Burrows97} Burrows, A., Marley, M.,
Hubbard, W.B., Lunine, J.I., Guillot, T., Saumon, D., Freedman, R.,
Sudarsky, D., \& Sharp, C. A Nongray Theory of Extrasolar Giant Planets and Brown Dwarfs.
\apj {\bf 491}, 856-875 (1997).

\bibitem{Sudarsky00} Sudarsky, D., Burrows, A.,
\& Pinto, P. Albedo and Reflection Spectra of Extrasolar Giant Planets.
\apj {\bf 538}, 885-903 (2000).

\bibitem{dyudina} Dyudina, U.A., Sackett, P.D., Bayliss, D.D.R., Seager, S.,
Porco, C., Throop, H.B., \& Dones, L. Phase light curves for extrasolar Jupiters and Saturns.
astro-ph/0406390 (2004).

\bibitem{GuillotShowman02} Guillot, T. \&
Showman, A.P.
Evolution of ``51 Pegasus b-like'' planets.
\aap {\bf 385}, 156-165 (2002).

\bibitem{ShowmanGuillot02} Showman, A.P.
\& Guillot, T.
Atmospheric circulation and tides of ``51 Pegasus b-like'' planets.
\aap {\bf 385}, 166-180 (2002).

\bibitem{SWS} Seager, S.,
Whitney, B.A, \& Sasselov, D.D.
Photometric Light Curves and Polarization of Close-in Extrasolar Giant Planets.
\apj {\bf 540}, 504-520 (2000).

\bibitem{kark99} Karkoschka, E. Methane, Ammonia, and Temperature 
Measurements of the Jovian Planets and Titan from CCD-Spectrophotometry.
{\it Icarus} {\bf 133}, 134-146 (1999).   

\bibitem{Sudarsky03} Sudarsky, D., Burrows, A.,
\& Hubeny, I. Theoretical Spectra and Atmospheres of Extrasolar Giant Planets.
\apj {\bf 588}, 1121-1148 (2003).

\bibitem{UnwinShao00} Unwin, S.C. \& Shao, M. Space Interferometry Mission.
in {\it Interferometry in Optical Astronomy} (eds. P. J. Lena
\& A. Quirrenbach) 754-761 (2000).

\bibitem{perryman} Perryman, M.A.C. GAIA Spectroscopy: Science and Technology.
in ASP Conference Proceedings, Vol. 298
(ed. Ulisse Munari) p. 3 (ASP Conf. Series, 2003).

\bibitem{arnold} Arnold, I. \& Schneider, J.
The detectability of extrasolar planet surroundings I. Reflected-light
photometry of unresolved rings.
astro-ph/0403330 (2004).

\bibitem{mather} Mather, J.C. and Stockman, H.S. 
{\it Proc. SPIE} {\bf 4013}, 2-16 (2000).

\bibitem{Akeson00} Akeson, R.L., Swain, M. R.,
\& Colavita, M.M. Differential phase technique with the Keck Interferometer. in {\it Interferometry in
Optical Astronomy} (ed. P. J. Lena) {\it Proc. SPIE} {\bf 4006}, 321-327 (2000).

\bibitem{AkesonSwain00} Akeson, R.L. \&
Swain, M.R. Differential Phase Observations of Extrasolar Planets with the Keck Interferometer.
in {\it From Giant Planets to Cool Stars},
(ed. C.A. Griffith \& M.S. Marley) p. 300 (ASP Conference Series Vol. 212, 2000) 

\bibitem{Paresce01} Paresce, F. Scientific
Objectives of the VLTI Interferometer. {\it The Messenger} {\bf 104}, 5-7 (2001).

\bibitem{beuzit} Beuzit, J.-L., et al. The Planet Finder project for the VLT.
in {\it SF2A-2004: Semaine de l'Astrophysique Francaise} 
(eds. F. Combes, D. Barret, T. Contini, F. Meynadier \& L. Paganii)
p. 32 (EdP-Sciences, Conference Series, 2004).

\bibitem{macintosh} Macintosh, B. et al. Extreme Adaptive Optics Planet Imager (XAOPI).
in {\it Techniques and Instrumentation for Detection of Exoplanets.}
(ed. Coulter, D. R.) pp. 272-282 (Proceedings of the SPIE, Vol. 5170, 2003).

\bibitem{tamura} Tamura, Itoh, I., \& Oasa, Y. Searches for
Extrasolar Planets with the Subaru Telescope: Companions
and Free-Floaters. in {\it Proceedings of the IAU 8th Asian-Pacific
Regional Meeting: Volume I}, 
(ed. Satoru Ikeuchi, John Hearnshaw and Tomoyuki Hanawa) pp. 73-76
(San Francisco, Astronomical Society of the Pacific, Vol. 289, 2003).

\bibitem{Hinz00} Hinz, P.M., Angel, J.R.P., Woolf, N.J.,
Hoffman, W.F., \& McCarthy, D.W. BLINC: a testbed for nulling interferometry in the thermal infrared.
in {\it Interferometry in
Optical Astronomy} (ed. P.J. Lena) {\it Proc. SPIE} {\bf 4006}, 349 (2000).

\bibitem{Hinz01} Hinz, P.M. Nulling interferometry for studying 
other planetary systems: Techniques and observations.
PhD Thesis, The University
of Arizona (2001).

\bibitem{hawarden} Hawarden, T.G., Gilmozzi, R., \& Hainaut, O.
Using a 100-meter ELT (e.g., ``OWL") for
Extrasolar Planet and Extrasolar Life Detection.
in {\it Scientific Frontiers in Research on Extrasolar Planets}
ASP Conference Series, Vol. 294 (ed. Drake Deming and Sara Seager)
pp. 581-586 (San Francisco: ASP, 2003).

\bibitem{davison} Davison, W.B., Woolf, N.J., \& Angel, J.R.P.
Design and Analysis of 20-m track mounted and 30-m telescopes.
in {\it Future Giant Telescopes} (eds. Angel, J.R.P. \& Gilmozzi, R.)
Proceedings of the SPIE, Vol. 4840, pp. 533-540 (2003).

\bibitem{strom} Strom, S. Toward a Giant Segmented Mirror Telescope: A
Progress Report from AURA's New Initiatives Office. American
Astronomical Society Meeting 203, \# 24.04 (2003).

\bibitem{antar} Aristidi, E., Agabi, A., Vernin, J., Azouit, M., Martin, F., Ziad, A., \&  Fossat, E.
Antarctic site testing: First daytime seeing monitoring at Dome C.
\aap {\bf 406}, L19-L22 (2003).

\bibitem{glenn} Schneider, G. et al.
Domains of observability in the near-infrared with HST/NICMOS and
adaptive-optics augmented large ground-based telescopes.
unpublished manuscript at
\verb"http://nicmosis.as.arizona.edu:8000/REPORTS/NICMOS_AO_WHITEPAPER.html"
(2002).

\bibitem{shaoepic} Shao, M., Levine, B.M., Wallace, J.K., Serabyn, E., \& Liu, D.T.
The Visible Nulling Coronagraph--Progress Towards Mission and Technology Development.
{\it BAAS} {\bf 203}, \#03.04 (2003). 


\bibitem{Trauger00} Trauger, J. \etal
Eclipse, A Direct Imaging Investigation of Nearby Planetary Systems.
{\it BAAS} {\bf 197}, \# 49.07, p. 1486 (2000).

\bibitem{Trauger01} Trauger, J., Hull, A.B.,
\& Redding, D.A. Eclipse Test Bed for Very High Contrast Space Astronomy.
{\it BAAS} {\bf 199}, \# 86.04 p. 1431 (2001).

\bibitem{beichman} Beichman, C.A., Coulter, D.R., Lindensmith, C., \& Lawson, P.R.
Selected Mission Architectures For The Terrestrial Planet Finder (TPF): Large, Medium, and Small.
in {\it Future Research Direction and Visions for Astronomy.}
(ed. Dressler, Alan M.), Proceedings of the SPIE, Vol. 4835, pp. 115-121 (2002).

\bibitem{Matthews01} Matthews, J.M., Kuschnig, R., \& Shkolnik, E.
Ultraprecise photometry from space: exploring 
pulsations with the ``Humble Space Telescope."
in {\it Proceedings of the SOHO 10/GONG 2000 Workshop: Helio- and asteroseismology at the dawn of the millennium}
(ed. A. Wilson) ESA SP-464, Noordwijk: ESA Publications Division, pp. 385-390 (2001).

\bibitem{origins} Roadmap for the Office of Science {\it Origins} theme. NASA/JPL-400-1060 (2003);
NASA 2003 Strategic Plan. Document \# NP-2003-01-298-HQ (2003).


\bibitem{quote} Keats, J. On First Looking into Chapman's Homer. (1816).



\bibitem{racine} Racine, R., Walker, G.A.H., Nadeau, D.,
Doyon, R. \& Marois, C. Speckle Noise and the Detection of Faint Companions.
\pasp {\bf 111}, 587-594 (1999).



\end{thebibliography}
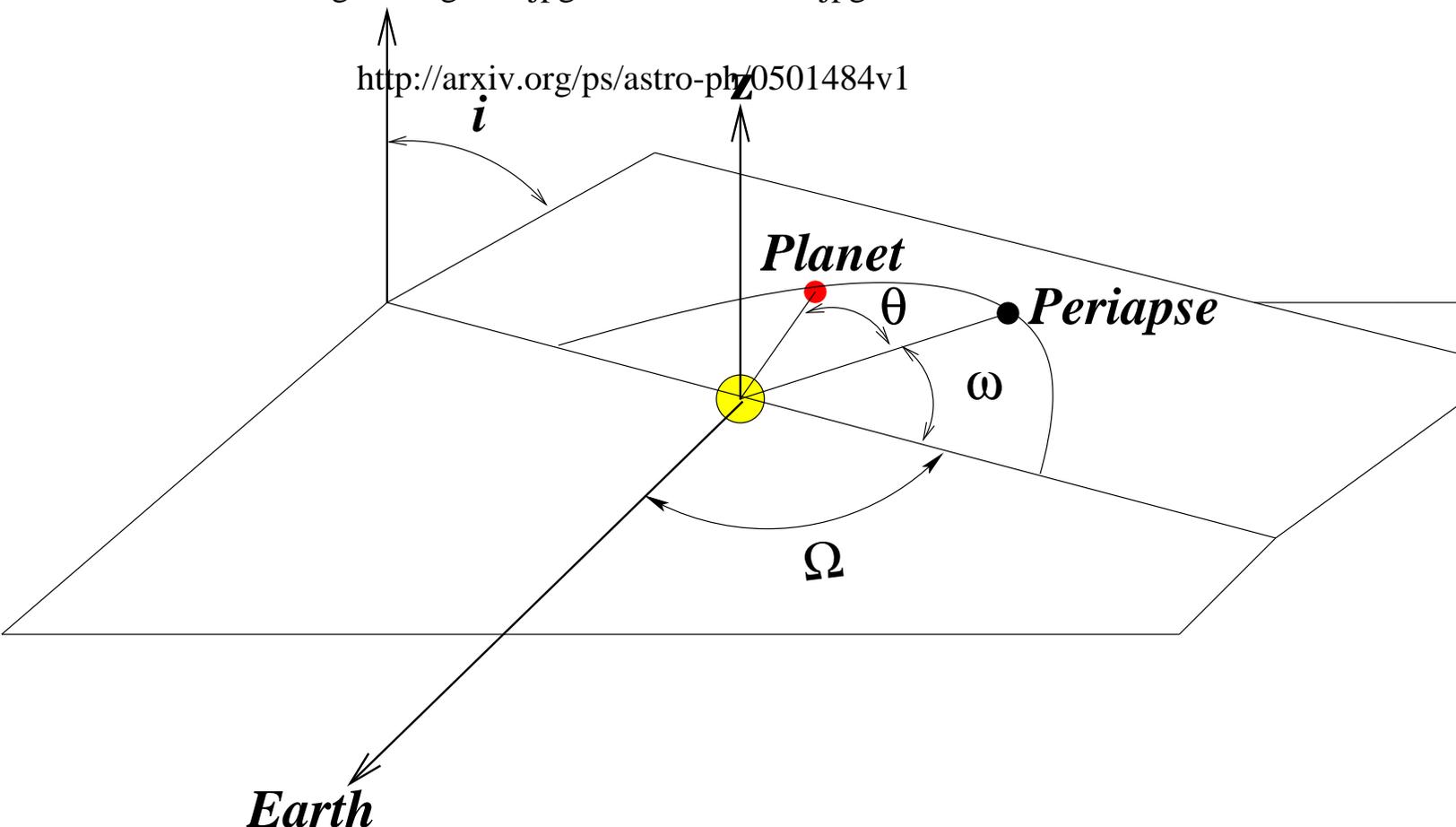
\end{document}